\documentclass[12pt]{article}
\usepackage{amsmath,amssymb,amsfonts,color,graphicx,cite,psfrag,soul,url}

\graphicspath{{figs/}}

\evensidemargin \oddsidemargin
\newcommand{\ttt}[1]{\texttt{#1}}
\newcommand{\snumentry}[2]{
\begin{minipage}[t]{1.2cm}\flushright\ttt{#1}\end{minipage}
\hspace{2mm}:
\begin{minipage}[t]{11cm}\noindent
#2\end{minipage}\\[1mm]
}
\newcommand{\numentry}[2]{
\begin{minipage}[t]{1.2cm}\flushright\texttt{#1}\end{minipage}
\hspace{2mm}:
\begin{minipage}[t]{11cm}\noindent
#2\end{minipage}\\[1mm]
}

\def\citeres#1{\mbox{Refs.~\cite{#1}}}

\def\citere#1{\mbox{Ref.~\cite{#1}}}

\allowdisplaybreaks

\begin{document}
\thispagestyle{empty}

\def\thefootnote{\fnsymbol{footnote}}

\begin{flushright}
DCPT/10/120, 
FR-PHENO-2010-024\\ 
IPPP/10/60, 
KA-TP-22-2010, 
KEK-TH-1379\\ 
MCnet/10/14, 
MPP-2010-94 
\end{flushright}

\vspace{0.5cm}

\begin{center}
{\large\sc {\bf 
Flavour Les Houches Accord: Interfacing Flavour related Codes}}

\vspace{2em}

{\sc 
F.~Mahmoudi$^{1,2}$%
\footnote{
email: mahmoudi@in2p3.fr
}%
, S.~Heinemeyer$^{3}$%
\footnote{
email: Sven.Heinemeyer@cern.ch
}%
, A.~Arbey$^{4}$%
, A.~Bharucha$^{5}$, \\
  T.~Goto$^{6}$%
, T. Hahn$^{7}$%
, U. Haisch$^{8}$%
, S.~Kraml$^{9}$%
, M.~Muhlleitner$^{10}$, \\
  J.~Reuter$^{11}$%
, P.~Skands$^{2}$%
, P.~Slavich$^{12}$
}
\vspace*{0.3cm}

{\sl 
$^1$Clermont Universit\'e, Universit\'e Blaise Pascal, CNRS/IN2P3,\\ 
LPC, BP 10448, F--63000 Clermont-Ferrand, France

\vspace*{0.1cm}
$^2$CERN Theory Division, Physics Department, CH-1211 Geneva 23, Switzerland

\vspace*{0.1cm}
$^3$Instituto de F\'isica de Cantabria (CSIC-UC), Santander, Spain
 
\vspace*{0.1cm}
$^4$Universit\'e de Lyon, France; Universit\'e Lyon 1, F--69622; CRAL, Observatoire de Lyon,\\ F--69561 Saint-Genis-Laval; CNRS, UMR 5574; ENS de Lyon, France

\vspace*{0.1cm}
$^5$IPPP, Department of Physics, University of Durham, Durham DH1 3LE, UK

\vspace*{0.1cm}
$^6$KEK Theory Center, Institute of Particle and Nuclear Studies,\\
KEK, Tsukuba, 305-0801 Japan

\vspace*{0.1cm}
$^7$Max-Planck-Institut f\"ur Physik, F\"ohringer Ring 6, D--80805 Munich, Germany

\vspace*{0.1cm}
$^8$Institut f\"ur Physik (WA THEP), Johannes Gutenberg-Universit\"at, \\
D--55099 Mainz, Germany

\vspace*{0.1cm}
$^{9}$Laboratoire de Physique Subatomique et de Cosmologie (LPSC), \\
UJF Grenoble 1, CNRS/IN2P3, 53 Avenue des Martyrs, 
F--38026 Grenoble, France

\vspace*{0.1cm}
$^{10}$Institut f\"ur Theoretische Physik, Karlsruhe Institute of Technology, \\
D--76128 Karlsruhe, Germany

\vspace*{0.1cm}
$^{11}$University of Freiburg, Institute of Physics, Hermann-Herder-Str. 3, \\
D--79104 Freiburg, Germany

\vspace*{0.1cm}
$^{12}$LPTHE, 4 Place Jussieu, F--75252 Paris, France
}
\end{center}
 
\vspace*{0.3cm}

\begin{abstract}
We present the Flavour Les Houches Accord (FLHA) which specifies a set of conventions for flavour-related parameters and observables. 
The FLHA uses the generic SUSY Les Houches Accord (SLHA) file structure. It defines the 
relevant Standard Model masses, Wilson coefficients, decay constants, bag parameters, flavour
observables, etc. The accord provides a universal and model-independent interface between codes evaluating and/or using flavour-related observables.  
\end{abstract}

\def\thefootnote{\arabic{footnote}}
\setcounter{page}{0}
\setcounter{footnote}{0}

\newpage


\section{Introduction}

In addition to the increasing number of refined approaches in the literature 
for calculating flavour-related observables, advanced programs dedicated to 
the calculation of such quantities, e.g. Wilson coefficients, branching ratios,
mixing amplitudes, renormalisation group equation (RGE) running including
flavour effects have recently been
developed~\cite{Mahmoudi:2007vz,Degrassi:2007kj,sufla,Rosiek:2010ug}. 
Flavour-related observables are also implemented by many other
non-dedicated public codes to provide additional checks for the models under
investigation 
\cite{Belanger:2008sj,Arbey:2009gu,Heinemeyer:1998yj,Lee:2003nta,Ellwanger:2005dv,Paige:2003mg,Porod:2003um,Djouadi:2002ze,Allanach:2001kg}.
The results are often subsequently used by other codes, e.g.\ as
constraints on the parameter space of the model under consideration
\cite{Lafaye:2004cn,Bechtle:2004pc,deAustri:2006pe,Master3}. 

At present, a small number of specialised interfaces between
the various codes exist. Such tailor-made interfaces are not easily generalised
and are time-consuming to construct and test for each specific
implementation. A universal interface would clearly be an advantage
here. 
Some time ago a similar problem arose in the context of Supersymmetry
(SUSY). The solution took the form of the SUSY Les Houches Accord
(SLHA)~\cite{Skands:2003cj,Allanach:2008qq}, which is nowadays frequently used
to exchange information between SUSY related codes, such as values for the soft SUSY-breaking 
parameters, particle masses and mixings, branching ratios etc.  
The SLHA has proved to be a robust solution, allowing information to be exchanged between 
different codes via ASCII files.
The detailed structure of these input and output files is described in
\citeres{Skands:2003cj,Allanach:2008qq}.  
While the first definitions, SLHA1~\cite{Skands:2003cj},
concerned the simplest version of the MSSM, the
SLHA2~\cite{Allanach:2008qq} also included definitions for the case of
complex parameters, $R$-parity violation, non-minimal flavour violation etc.

The goal of this article is to exploit the existing organisational structure
of the SLHA and use it to define an accord for the exchange of flavour
related quantities, which we refer to as the ``Flavour Les Houches Accord''
(FLHA). In brief, the purpose of this Accord is thus to
present a set of generic 
definitions for an input/output file structure which provides a
universal framework for interfacing flavour-related programs. Furthermore, the
standardised format will provide the users with a clear and well-structured
result that could eventually be used for other purposes. 
We stress that in cases of ambiguity in the precise definition of a given quantity, it is the 
responsibility of the authors of the specific program to document in detail the definition they use.

The structure is set up in such a way that the SLHA and the FLHA can be
used together or independently.  
Obviously, some of the SLHA entries, such as measured parameters in the
Standard Model (SM) and the Cabibbo-Kobayashi-Maskawa (CKM) matrix
elements, are also needed for flavour observable 
calculations. Therefore, a FLHA file can indeed contain a SLHA block if
necessary. For this reason, and also for the sake of clarity, the
new FLHA block
names start with ``\texttt{F}''. Also, in order to avoid any confusion,
the SLHA blocks are not modified or redefined in the FLHA.
Instead, if a block needs to be extended to meet the requirements of flavour physics, a
new ``\texttt{F}'' block is defined.  

Note that different codes may \emph{technically} achieve the FLHA
input/output in different ways. The details of how to
``switch on'' the FLHA input/output for a particular program should be
described in the manual of that program and are not covered here. 
For the SLHA, libraries have been developed to permit an easy
implementation of the input/output routines~\cite{Hahn:2006nq}. In
principle these programs could be extended to include the FLHA as well. 

It should be noted that, while the SLHA was developed especially for the
case of SUSY, the FLHA is, at least in principle, model
independent. While it is possible to indicate the choice of model in a
specific block, the general structure used for the information exchange can
be applied to any model.


\section{Definitions of the interfaces}

The FLHA input and output files are described in this section. 


\subsection{General structure}

Following the general structure for the SLHA~\cite{Skands:2003cj,Allanach:2008qq} we assume
the following: 
 
\begin{itemize}

\item All quantities with dimensions of energy (mass) are in GeV (GeV$/c^2$). 

\item Particles are identified by their PDG particle codes. See Appendix
  \ref{app:pdg} for lists of these, relevant for flavour observables. 

\item The first character of every line is reserved for control and
  comment statements. The first character of data lines should be empty. 

\item In general, a formatted output should be used for write-out, to
  avoid ``messy-looking'' files, while a free format should be used for
  read-in, to avoid misalignment etc.~leading to program crashes.  

\item Read-in should be performed in a case-insensitive way, again to
  increase stability.  

\item The general format for all real numbers is the FORTRAN format
 E16.8\footnote{E16.8:  
  a 16-character wide real number in scientific notation, whereof
  8 digits are decimals, e.g., ``\texttt{-0.12345678E+000}''.}. 
  The large number of digits is used to avoid any possible numerical
  precision issue, and since it is no more difficult for e.g. the spectrum
  calculator to write out such a number than a shorter version. For typed
  input, this merely means that at least 16 spaces are reserved for the number,
  but e.g. the number \texttt{123.456} may be typed in ``as is''. See
 also the example file in Appendix \ref{app:example}.

\item A ``\texttt{\#}'' 
mark anywhere means that the rest of the line is intended as a comment and
should be ignored by the reading program.  

\item Any input and output is divided into sections in the form of
  ``blocks''.  

\item To clearly identify the blocks of the FLHA, the first letter of
  the name of a block is an ``\texttt{F}''.
  There are two exceptions to this rule: blocks borrowed from the
    SLHA, which keep their original name, and blocks containing
    imaginary parts, which start with ``\texttt{IMF}''.

\item A ``\texttt{BLOCK Fxxxx}''  
(with the ``\texttt{B}'' being the first
character on the line) marks the beginning of entries belonging to 
the block named ``\texttt{Fxxxx}''. For instance, 
``\texttt{BLOCK FMASS}'' marks that all following lines until the next
``\texttt{BLOCK}'' statement contain mass values, to be read 
in a specific format, intrinsic to the \texttt{FMASS} block.

\item The FLHA is designed to be compatible with SLHA (and with any 
  future accords that follow the same general
  structure). For generic models, e.g. ones for
  which an accord does not yet exist, we provide a fall-back solution
  by generalising the SLHA block \texttt{MODSEL}. This takes the form of a new
  FLHA-specific block, \texttt{FMODSEL}, 
  with entries as described below. For models for which a
  more specific accord does exist, e.g. SLHA, a complete model of that
  type can be specified instead, such that \texttt{FMODSEL} is redundant
  and should in most cases be absent. In case it is present and there are 
  conflicts, the most specific accord takes precedence, such that internal
  consistency of the accord with the smallest scope is guaranteed.
  In the case of \texttt{MODSEL} vs.\ \texttt{FMODSEL}, \texttt{MODSEL} 
  would thus take precedence, however it would be prudent  
  to issue warnings if both are present, especially if their
  contents differ.  
  Also note that although there is the possibility to reuse more specific accords, one is not forced to do so. Thus, even for a SUSY
  model, an FLHA user who does not wish to deal with an entire SLHA 
  spectrum could just use \texttt{FMODSEL}. In 
  that case, the file would be treated just like any other FLHA file, i.e.
  without any reference to the SUSY-specific parts of the SLHA. The
  only difference from a practical perspective would be that
  SLHA-specific tools would then not be able to process the file
  correctly.

\item The order of the blocks is arbitrary, although it is in general
  good practice to put \texttt{FMODSEL} or \texttt{MODSEL} near the beginning.

\end{itemize}
Further definitions can be found in section~3 of \citere{Skands:2003cj}.\\
\\
The following general structure for the FLHA file is proposed:
\begin{itemize}

\item \texttt{BLOCK FCINFO}:
Information about the flavour code used to prepare the FLHA file.

\item \texttt{BLOCK FMODSEL}:
Basic information about the underlying model used for the calculations, for
generic models. 
This is the only place where ``model dependent'' information can be
found. In the case of SUSY models with complete SLHA spectra, 
the SLHA \texttt{BLOCK MODSEL} is used instead and overrides \texttt{FMODSEL}
if both are present.

\item \texttt{BLOCK SMINPUTS}: 
Measured values of SM parameters used for the calculations.

\item \texttt{BLOCK VCKMIN}: 
Input parameters of the CKM matrix in the Wolfenstein parameterisation.

\item \texttt{BLOCK UPMNSIN}: 
Input parameters of the PMNS neutrino mixing matrix in the PDG parameterisation.

\item \texttt{BLOCK VCKM}: 
Real part of the CKM matrix elements.

\item \texttt{BLOCK IMVCKM}: 
Imaginary part of the CKM matrix elements.

\item \texttt{BLOCK UPMNS}: 
Real part of the PMNS matrix elements.

\item \texttt{BLOCK IMUPMNS}: 
Imaginary part of the PMNS matrix elements.

\item \texttt{BLOCK FMASS}:
Masses of quarks, mesons, hadrons, etc.

\item \texttt{BLOCK FLIFE}:
Lifetime (in seconds) of mesons, hadrons, etc.

\item \texttt{BLOCK FCONST}:
Decay constants.

\item \texttt{BLOCK FCONSTRATIO}:
Ratios of decay constants.

\item \texttt{BLOCK FBAG}:
Bag parameters.

\item \texttt{BLOCK FWCOEF}: 
Real part of the Wilson coefficients.

\item \texttt{BLOCK IMFWCOEF}: 
Imaginary part of the Wilson coefficients.

\item \texttt{BLOCK FOBS}:
Prediction of flavour observables.

\item \texttt{BLOCK FOBSERR}:
Theory error on the prediction of flavour observables.

\item \texttt{BLOCK FOBSSM}:
SM prediction for flavour observables.

\item \texttt{BLOCK FDIPOLE}:
Prediction of electric and magnetic dipole moments.

\item \texttt{BLOCK FPARAM}: 
Process dependent variables and parameters.

\end{itemize}
More details on each block are given in the following.


\subsection{Definition of the blocks}
The FLHA input and output blocks are described in the following.

\subsection*{\texttt{BLOCK FCINFO}}

Flavour code information, including the name and the version of the program:\\
\numentry{1}{Name of the flavour calculator}
\numentry{2}{Version number of the flavour calculator}
  Optional warning or error messages can also be specified:\\
\numentry{3}{If this entry is present, it means warnings were produced by the
  flavour calculator. The resulting file may still be used. The entry
  should contain a description of the problem (string).}   
\numentry{4}{If this entry is present, it means errors were produced by the
  flavour calculator. The resulting file should not be used. The entry
  should contain a description of the problem (string).}  
A string is as usual a sequence of ASCII characters. 
This block is purely informative, and is similar to \texttt{BLOCK
  SPINFO} in the SLHA.  


\subsection*{\texttt{BLOCK MODSEL}}

This block is part of the SLHA. Its presence in an FLHA file signals
that the file contains, in addition to the FLHA-specific blocks, a
complete and self-consistent SLHA spectrum, for which \texttt{MODSEL} defines
the switches and options used. It is defined as in the SLHA2. For non-SLHA models, see instead \texttt{BLOCK FMODSEL} below. Note that, if the user does not wish to
  provide a complete set of SLHA blocks, \texttt{MODSEL} should not be
  used. Instead, see \texttt{BLOCK FMODSEL} below.


\subsection*{\texttt{BLOCK FMODSEL}}

In the case of non-SLHA models this block provides switches and
options for model selection. It is similar to the  
SLHA2 \texttt{BLOCK MODSEL}, but extended to allow more
flexibility. 
It is advised that one only uses \texttt{BLOCK MODSEL} for SLHA models and \texttt{BLOCK FMODSEL} for non-SLHA models, to avoid double definitions. 

\numentry{1}{Choice of model. By default, a
minimal type of model will always be assumed. Possible
values are (note: when giving a complete SLHA spectrum, use
\texttt{BLOCK MODSEL}  instead. The options here are only intended to
cover the cases when a complete SLHA spectrum is not provided.):\\[2mm]
\snumentry{-1}{SM}
\snumentry{0}{General MSSM simulation} 
\snumentry{1}{(m)SUGRA model} 
\snumentry{2}{(m)GMSB model}
\snumentry{3}{(m)AMSB model}
\snumentry{4}{...}
\snumentry{3O}{General THDM}
\snumentry{31}{THDM type I}
\snumentry{32}{THDM type II}
\snumentry{33}{THDM type III}
\snumentry{34}{THDM type IV}
\snumentry{35}{...}
\snumentry{99}{other model. This choice requires a string given in the
  entry \ttt{99}}
}\\
\numentry{5}{(Default=\ttt{0}) CP violation. Switches defined are:\\
\snumentry{0}{ CP violation is completely neglected. 
No information on the CKM phase is used.}
\snumentry{1}{CP is violated, but only by the standard CKM
phase. All other phases are assumed zero.}
\snumentry{2}{CP is violated. Completely general CP phases
allowed.}
}\\
\numentry{6}{(Default=\ttt{0}) Flavour violation. Switches defined are:\\
\snumentry{0}{No flavour violation. 
}
\snumentry{1}{Quark flavour is violated.}
\snumentry{2}{Lepton flavour is violated.}
\snumentry{3}{Lepton and quark flavour is violated.}
}
\numentry{99}{a string that defines other models is used only if
  entry~\ttt{1} is given as~\ttt{99}, otherwise it is ignored.}

The definition of the different Two-Higgs Doublet Model (THDM)
  types is given in Appendix~\ref{app:2hdm}.

\smallskip
Private blocks can also be constructed by the user, 
for instance \ttt{BLOCK MYMODEL}, to contain parameters specific to other models. It is advised that all such relevant model
parameters are provided in this way. Due to the user-defined specific
structure of these blocks it is not required that they are universally 
recognised.


\subsection*{\texttt{BLOCK SMINPUTS}}

In general, the spectrum of the SM particles plays a crucial role in flavour physics.  
Consequently, experimental measurements of masses and coupling constants
at the electroweak scale are required. 
The block containing these quantities in the SLHA is
\texttt{SMINPUTS}. We borrow this block from SLHA as it is, and reproduce it  here for completeness. 

It is also important to note that experimental results for all quantities available at present, e.g. $\alpha_s$ or the running bottom quark mass, are clearly obtained based on 
the assumption that the SM is the underlying theory. Extending the field content of the SM to that of a New Physics (NP) Model means that the \emph{same} measured results would be obtained
for \emph{different} values of these quantities.  However, since these
values are not known, all parameters contained in the block
\texttt{SMINPUTS} should be the ``ordinary'' ones obtained from SM fits,
i.e.\ with no NP corrections included. Any flavour code itself is then
assumed to convert these parameters into ones appropriate to an NP
framework. 

It should be noted that some programs have
hard-coded defaults for some of these parameters, hence only a subset
may sometimes be available as free inputs. The parameters are:\\ 
\numentry{1}{$\alpha_\mathrm{em}^{-1}(m_{Z})^{\overline{\mathrm{MS}}}$, 
  inverse electromagnetic coupling at the $Z$ pole in the
  $\overline{\mathrm{MS}}$ scheme  (with 5 active flavours).}  
\numentry{2}{$G_F$, Fermi constant (in units of GeV$^{-2}$).}
\numentry{3}{$\alpha_s(m_{Z})^{\overline{\mathrm{MS}}}$, strong coupling
  at the $Z$ pole in the $\overline{\mathrm{MS}}$ scheme (with 5 active
  flavours).}%
\numentry{4}{$m_Z$, pole mass.}
\numentry{5}{$m_b(m_b)^{\overline{\mathrm{MS}}}$, bottom quark running mass
  in the $\overline{\mathrm{MS}}$ scheme (with 5 active flavours).} 
\numentry{6}{$m_t$, top quark pole mass.}
\numentry{7}{$m_\tau$, tau pole mass.}


\vspace*{-1.2cm}
\subsection*{\texttt{BLOCK VCKMIN}}
This block is strictly identical to the SLHA2 \texttt{BLOCK VCKMIN}.
The parameters are:\\[2mm]
\numentry{1}{$\lambda$}
\numentry{2}{$A$}
\numentry{3}{$\bar{\rho}$}
\numentry{4}{$\bar{\eta}$}
We use the PDG definition, Eq.~(11.4) of \citere{Nakamura:2010zzi}, which
is exact to all orders in $\lambda$, and also the PDG parameterisation for the phase convention. For the output we use the SLHA2 blocks \texttt{VCKM} and \texttt{IMVCKM} which do not rely on any specific convention. 


\subsection*{\texttt{BLOCK UPMNSIN}}
This block is strictly identical to the SLHA2 \texttt{BLOCK UPMNSIN}.
The parameters are:\\[2mm]
\numentry{1}{$\theta_{12}$}
\numentry{2}{$\theta_{23}$}
\numentry{3}{$\theta_{13}$}
\numentry{4}{$\delta$}
\numentry{5}{$\alpha_1$}
\numentry{6}{$\alpha_2$}
We use the PDG parameterisation, Eq.~(13.30) of \citere{Nakamura:2010zzi}.
All the angles and phases should be given in radians. For the output we use \texttt{BLOCK UPMNS} and \texttt{BLOCK IMUPMNS}, exactly in the same way as in SLHA2. 


\subsection*{\texttt{BLOCK FMASS}}

The block \texttt{BLOCK FMASS} contains the mass spectrum for the
particles involved, in addition to the 
SLHA \texttt{BLOCK MASS} which only contains pole masses and to the
SLHA \ttt{BLOCK SMINPUTS} which contains quark masses.
If a mass is given in two blocks the block \ttt{FMASS} overrules the
other blocks.
In \ttt{FMASS} we
specify additional information concerning the renormalisation scheme as
well as the scale at which the masses are given and thus allow for
larger flexibility. The standard for each
line in the block should correspond to the FORTRAN format, 
\begin{center} 
\texttt{(1x,I9,3x,1P,E16.8,0P,3x,I2,3x,1P,E16.8,0P,3x,'\#',1x,A)},
\end{center} 
where the first nine-digit integer should be the PDG code of a particle,
followed by a double precision number for its mass. The next integer
corresponds to the renormalisation scheme, and finally the last double
precision number indicates the energy scale (0 if not relevant). 
An additional comment could be given after \texttt{\#}.
\newpage
The schemes are defined as follows:\\[2mm]
\numentry{0}{pole}
\numentry{1}{$\overline{\mathrm{MS}}$}
\numentry{2}{$\overline{\mathrm{DR}}$}
\numentry{3}{1S}
\numentry{4}{kin}
\numentry{5}{\ldots}

For the definition of $\overline{\mathrm{DR}}$ scheme see
  \cite{Siegel:1979wq,Jack:1994rk}. 


\subsection*{\texttt{BLOCK FLIFE}}

The block \texttt{BLOCK FLIFE} contains the lifetimes, in seconds, of mesons and hadrons. The standard for each line
in the block should correspond to the FORTRAN format 
\begin{center} 
\texttt{(1x,I9,3x,1P,E16.8,0P,3x,'\#',1x,A)},
\end{center} 
where the first nine-digit integer should be the PDG code of a particle
and the double precision number its lifetime.  


\subsection*{\texttt{BLOCK FCONST}}

The block \texttt{BLOCK FCONST} contains the decay constants in
GeV. The renormalisation scheme and scale are also specified to allow for
  large flexibility. The standard for each line in the block should
correspond to the FORTRAN format  
\begin{center} 
\texttt{(1x,I9,3x,I2,3x,1P,E16.8,0P,3x,I2,3x,1P,E16.8,0P,3x,'\#',1x,A)},
\end{center} 
where the first nine-digit integer should be the PDG code of a particle,
the second integer the number associated with the decay constant, the
double precision number the value of this decay constant, 
the following integer stands for the renormalisation scheme
  (see below) and finally the last double precision number indicates the
  renormalisation scale $Q$ (0 in the case of renormalisation group
  invariant parameters). An additional comment could be given after
  \texttt{\#}. 

The decay constants for the mesons that are used most often, which have several decay constants associated with them, are defined as:\\[0.3cm]
\numentry{321}{$K^+$\\
\numentry{1}{$f_K$ in GeV}
\numentry{11}{$h_K$ in GeV$^3$}
}
\numentry{221}{$\eta$\\
\numentry{1}{$f_{\eta}^q$ in GeV}
\numentry{2}{$f_{\eta}^s$ in GeV}
\numentry{11}{$h_{\eta}^q$ in GeV$^3$}
\numentry{12}{$h_{\eta}^s$ in GeV$^3$}
}
\numentry{213}{$\rho(770)^+$\\
\numentry{1}{$f_{\rho}$ in GeV}
\numentry{11}{$f^T_{\rho}$ in GeV}
}
\numentry{223}{$\omega(782)$\\
\numentry{1}{$f_{\rho}^{q}$ in GeV}
\numentry{2}{$f_{\rho}^{s}$ in GeV}
\numentry{11}{$f^{T,q}_{\rho}$ in GeV}
\numentry{12}{$f^{T,s}_{\rho}$ in GeV}
}
More details, and definitions for the decay constants ($f$, $h$, etc.) can be found in Appendix~\ref{app:decayconst}. 

Finally, the schemes are defined as follows:\\[2mm]
\numentry{0}{renormalisation group invariant (with renormalisation scale set to 0)}
\numentry{1}{$\overline{\mathrm{MS}}$}
\numentry{2}{MOM}
\numentry{3}{SMOM}
\numentry{4}{\ldots}


\subsection*{\texttt{BLOCK FCONSTRATIO}}

The block \texttt{BLOCK FCONSTRATIO} contains the ratios of decay constants, which often have less uncertainty than the decay constants themselves. The
ratios are specified by the two PDG codes in the form
f(code1)/f(code2). The standard for each line in the block should
correspond to the FORTRAN format 
\begin{center}
\texttt{(1x,I9,3x,I9,3x,I2,3x,I2,3x,1P,E16.8,0P,3x,I2,3x,1P,E16.8,0P,3x,'\#',1x,A)},
\end{center}
where the two nine-digit integers should be the two PDG codes of
particles, the third and fourth integers the numbers of the decay constants, which
correspond to the second index of the entry in \texttt{BLOCK FCONST}, the double precision number the ratio of the decay constants and the following integer stands for the renormalisation scheme and finally the last double precision number indicates the renormalisation scale $Q$ in the same way as in \texttt{BLOCK FCONST}.   


\subsection*{\texttt{BLOCK FBAG}}

The block \texttt{BLOCK FBAG} contains the bag parameters. The
renormalisation scheme as well as the scale at which the bag parameters
are evaluated are also specified to allow for large flexibility. The
standard for each line in the block should correspond to the FORTRAN
format 
\begin{center}
\texttt{(1x,I9,3x,I2,3x,1P,E16.8,0P,3x,I2,3x,1P,E16.8,0P,3x,'\#',1x,A)},
\end{center}
where the first nine-digit integer should be the PDG code of a particle, the
second integer the number associated with the bag parameter, the double
precision number the value of the bag parameter, the following integer
stands for the renormalisation scheme and finally the last double
precision number indicates the renormalisation scale $Q$ (0
in the case of renormalisation group invariant bag parameters).   
An additional comment could be given after \texttt{\#}.

Bag parameters for $B_d^0-\bar{B}_d^0$ (and similarly for $B_s^0-\bar{B}_s^0$, $K^0-\bar{K}^0$ and $D^0-\bar{D}^0$) mixing are listed as:
\\[2mm]
\numentry{511}{$B_d^0$\\
\numentry{1}{$B_1(Q)$}
\numentry{2}{$B_2(Q)$}
\numentry{3}{$B_3(Q)$}
\numentry{4}{$B_4(Q)$}
\numentry{5}{$B_5(Q)$}
}
The schemes are defined in the same way as in \texttt{BLOCK
    FCONST} and the Bag parameter definitions and normalisations can be
found in Appendix~\ref{app:bagpar}. 


\subsection*{\texttt{BLOCK FWCOEF Q= \ldots}}

The block \texttt{BLOCK FWCOEF Q= \ldots} contains the real part of the
Wilson coefficients at the scale \texttt{Q}.
The Wilson coefficients are calculated for the required set of
operators. A list of the most relevant effective operators is given in
Appendix~\ref{app:operators} and we recommend the user to use the
normalisation as given in that appendix (the detailed information about the scheme
and the normalisation of the effective interactions has to be provided in the program manual, and a short comment can be given in the FLHA file). 

Note that there can be several such blocks for different scales \texttt{Q}. 

The coefficients $C^{(k)}_{i,j}$ at order $k$ in the perturbative expansion in $\alpha(\mu)$ ($j=e$) or $\alpha_s(\mu)$ ($j=s$), have to be specified individually.  The following convention for the perturbative expansion is used: 
\begin{eqnarray}
  C_{i}(\mu) &=&
  C^{(0)}_{i}(\mu)
  + \dfrac{\alpha_s(\mu)}{4\pi} C^{(1)}_{i,s}(\mu)
  + \left( \dfrac{\alpha_s(\mu)}{4\pi} \right)^2 C^{(2)}_{i,s}(\mu)
\nonumber\\&&
  + \dfrac{\alpha(\mu)}{4\pi} C^{(1)}_{i,e}(\mu)
  + \dfrac{\alpha(\mu)}{4\pi}
    \dfrac{\alpha_s(\mu)}{4\pi} C^{(2)}_{i,es}(\mu)
  + \cdots \;.
\label{eq:WCexpansion}
\end{eqnarray}
Therefore the couplings should not be included in the Wilson coefficients.     

The first two entries on each line in \texttt{BLOCK FWCOEF}
should consist of two integers 
defining the fermion structure of the operator and the operator
structure itself. These two numbers do not describe the operator fully,
including normalisation etc.\ , but 
together act as a unique identifier for any possible Wilson coefficient. 
Consequently, the user has to take care that a consistent
normalisation including prefactors etc. is used in specifying the Wilson
coefficients. 
The most relevant operators are
listed in Appendix~\ref{app:operators}. 
As an example, for the operator $O_1$,
\begin{equation}
O_1 = (\bar{s} \gamma_{\mu} T^a P_L c)
       (\bar{c} \gamma^{\mu} T^a P_L b) \;,
\label{O1}
\end{equation}
the definition of the two numbers is given as follows. 
The fermions are encoded by their PDG code in two-digit form, which is the same for particles and
antiparticles, as given in Table~\ref{tab:pdgcodes}. 
Correspondingly, the first integer number defining $O_1$, containing the
fermions $\bar s c \bar c b$, is given by~03040405.
The various operator sub-structures are defined in Table~\ref{tab:opcodes}. 
Correspondingly, the second integer number defining $O_1$, containing
the operator structure $\gamma_\mu T^a P_L \gamma^\mu T^a P_L$ is given by~6161.

A few more rules are needed for an unambiguous definition.
\begin{itemize}
\item
If part of an operator appears without fermions (as it is possible, e.g., for
$F_{\mu\nu}$) it should appear right-most, so that the encoded fermions
correspond to the left-most part of the operator.
\item
In the case of a possible ambiguity, for instance
$O_1 = (\bar s \gamma_\mu T^a P_L c) (\bar c \gamma^\mu T^a P_L b)$
corresponding to 03040405~6161 and
$O_1 =  (\bar c \gamma_\mu T^a P_L b) (\bar s \gamma^\mu T^a P_L c)$
corresponding to 04050304~6161,
the ``smaller'' number, i.e.\ in this case 03040405~6161, should be
used.\\[-1.5em] 
\end{itemize}

\begin{table}[!t]
\begin{center}
\begin{tabular}{|c|c|c||c|c|c|}
\hline
name & PDG code & two-digit number & 
name & PDG code & two-digit number \\
\hline
    $d$ & 1 & 01 & $e$     & 11 & 11 \\
    $u$ & 2 & 02 & $\nu_e$ & 12 & 12 \\
    $s$ & 3 & 03 & $\mu$   & 13 & 13 \\  
    $c$ & 4 & 04 & $\nu_\mu$ & 14 & 14 \\
    $b$ & 5 & 05 & $\tau$    & 15 & 15 \\
    $t$ & 6 & 06 & $\nu_\tau$ & 16 & 16 \\
    $\sum_q q$ &  & 07 & $\sum_l l$ &   & 17\\ 
    $\sum_q q Q_q$ & & 08 & $\sum_l l Q_l$ & & 18\\
\hline
\end{tabular}
\caption{PDG codes and two-digit number identifiers of quarks and
  leptons. The summations are over active fermions.\label{tab:pdgcodes}}
\end{center}
\end{table}
\begin{table}[!t]
\begin{center}
\begin{tabular}{|c|c||c|c||c|c|}
\hline
operator & number & 
operator & number & 
operator & number \\
\hline
    $1$                   & 30 & $T^a$    & 50 & $\delta_{ij}$     & 70 \\
    $P_L$                 & 31 & $P_L T^a$ & 51 & $P_L \delta_{ij}$ & 71 \\
    $P_R$                 & 32 & $P_R T^a$ & 52 & $P_R \delta_{ij}$ & 72 \\
    $\gamma^\mu$          & 33 & $\gamma^\mu T^a$ & 53 &
                                 $\gamma^\mu \delta_{ij}$ & 73 \\
    $\gamma_5$            & 34 & $\gamma_5 T^a$ & 54 &
                                 $\gamma_5 \delta_{ij}$ & 74 \\
    $\sigma^{\mu\nu}$     & 35 & $\sigma^{\mu\nu} T^a$ & 55 &
                                $\sigma^{\mu\nu} \delta_{ij}$ & 75 \\
    $\gamma^\mu \gamma^\nu \gamma^\rho$ & 36 &  
    $\gamma^\mu \gamma^\nu \gamma^\rho T^a$ & 56 &
    $\gamma^\mu \gamma^\nu \gamma^\rho \delta_{ij}$ & 76 \\
    $\gamma^\mu \gamma_5$ & 37 &
    $\gamma^\mu \gamma_5 T^a$ & 57 &
     $\gamma^\mu \gamma_5 \delta_{ij}$ & 77 \\
    $\gamma^\mu P_L$      & 41 & $\gamma^\mu T^a P_L$ & 61 &
                                 $\gamma^\mu \delta_{ij} P_L$ & 81 \\  
    $\gamma^\mu P_R$      & 42 & $\gamma^\mu T^a P_R$ & 62 &
                                 $\gamma^\mu \delta_{ij} P_R$ & 82 \\
    $\sigma^{\mu\nu} P_L$ & 43 & $\sigma^{\mu\nu} T^a P_L$ & 63 &
                                $\sigma^{\mu\nu} \delta_{ij} P_L$ & 83 \\
    $\sigma^{\mu\nu} P_R$ & 44 & $\sigma^{\mu\nu} T^a P_R$ & 64 &
                                $\sigma^{\mu\nu} \delta_{ij} P_R$ & 84 \\
    $\gamma^\mu \gamma^\nu \gamma^\rho P_L$ & 45 &
    $\gamma^\mu \gamma^\nu \gamma^\rho T^a P_L$ & 65 &
    $\gamma^\mu \gamma^\nu \gamma^\rho \delta_{ij} P_L$ & 85 \\
    $\gamma^\mu \gamma^\nu \gamma^\rho P_R$ & 46 &
    $\gamma^\mu \gamma^\nu \gamma^\rho T^a P_R$ & 66 &
    $\gamma^\mu \gamma^\nu \gamma^\rho \delta_{ij} P_R$ & 86 \\
    $F_{\mu\nu}$          & 22 & $G_{\mu\nu}^a$ & 21 & & \\ 
\hline
\end{tabular}
\caption{Two-digit number definitions for the operators.
$T^a$ ($a = 1 \ldots 8$) denote the $SU(3)_C$ generators,
$P_{L,R} = \frac{1}{2} (1 \mp \gamma_5)$, and
$(T^a)_{ij} (T^a)_{kl} = \frac{1}{2} (\delta_{il} \delta_{kj} 
                      - 1/N_c \, \delta_{ij} \delta_{kl})$, where
                      $i,j,k,l$ are colour indices. 
\label{tab:opcodes}}
\end{center}
\end{table}

The third entry specifies the order of the perturbative expansion in Eq.~(\ref{eq:WCexpansion}). The information about the order is given by a two-digit number $xy$, where $x$ indicates $\mathcal{O}(\alpha^x)$ and $y$ indicates $\mathcal{O}(\alpha_s^y)$, and 0 indicates $C_i^{(0)}$,  e.g.:\\[2mm]
\numentry{00}{$C^{(0)}_{i}(Q)$}
\numentry{01}{$C^{(1)}_{i,s}(Q)$}
\numentry{02}{$C^{(2)}_{i,s}(Q)$}
\numentry{10}{$C^{(1)}_{i,e}(Q)$}
\numentry{11}{$C^{(2)}_{i,es}(Q)$}
\numentry{99}{total}
The Wilson coefficients can be provided either via separate New Physics and SM contributions, or as a total contribution of both New Physics and SM, depending on the code generating them. To avoid any confusion, the fourth entry must specify whether the given
Wilson coefficients correspond to the SM contributions, New Physics contributions or to the sum of them, using the following definitions:\\[2mm]
\numentry{0}{SM}
\numentry{1}{NP}
\numentry{2}{SM+NP}
The New Physics model is the model specified in the \texttt{BLOCK FMODSEL}.

\noindent The standard for each line in the block should thus correspond to the
FORTRAN format
\begin{center}
\texttt{(1x,I8,1x,I4,3x,I2,3x,I1,3x,1P,E16.8,0P,3x,'\#',1x,A)},
\end{center}
where the eight-digit integer specifies the fermion content, the
four-digit integer the operator structure, the two-digit integer the
order at which the Wilson coefficients are calculated followed by the
one-digit integer specifying the model, and finally the double precision
number gives the real part of the Wilson coefficient.


\subsection*{\texttt{BLOCK IMFWCOEF Q= \ldots}}

The block \texttt{BLOCK IMFWCOEF} contains the imaginary part of the
Wilson coefficients at the scale \texttt{Q}. 
The structure is exactly the same as that of the \texttt{BLOCK FWCOEF}.


\subsection*{\texttt{BLOCK FOBS}}

The block \texttt{BLOCK FOBS} contains the predictions for the flavour observables. The structure of this block is based on the decay
table in SLHA format. The decay is defined by the PDG number of the
parent, the type of observable, the value of the observable, the energy dependence $q$ for the observable (0 if not relevant), number of daughters and PDG IDs of the daughters.\\ 
The types of the observables are defined as follows:\\[2mm]
\numentry{1}{Branching ratio}
\numentry{2}{Ratio of the branching ratio to the SM value}
\numentry{3}{Asymmetry -- CP}
\numentry{4}{Asymmetry -- isospin}
\numentry{5}{Asymmetry -- forward-backward}
\numentry{6}{Asymmetry -- lepton-flavour}
\numentry{7}{Mixing}
\numentry{8}{\ldots}
The standard for each line in the block should correspond to the FORTRAN format
\begin{center}
\texttt{(1x,I9,3x,I2,3x,1P,E16.8,0P,3x,1P,E16.8,0P,3x,I1,3x,I9,3x,I9,3x,\ldots,3x,'\#',1x,A)},
\end{center}
where the first nine-digit integer should be the PDG code of the parent
decaying particle, the second integer the type of the observable, the
double precision number the value of the observable, the next double precision number the energy scale $q$, the next integer the number of daughters, and the following nine-digit integers the PDG codes of the daughters. It is strongly advised to give the descriptive
name of the observable as a comment. More details and definitions regarding the meson mixing are provided in Appendix~\ref{app:mixing}.

For more specific conventions for each observable, the user is encouraged to refer to the manual of the flavour calculator. 


\subsection*{\texttt{BLOCK FOBSERR}}

The block \texttt{BLOCK FOBSERR} contains the theoretical error for
flavour observables, with a similar structure to that of
\texttt{BLOCK FOBS}, but with the double precision number for the value of
the observable replaced by two double precision numbers for the minus
and plus uncertainties. 

In a similar way, for every block \texttt{Fname}, a corresponding block containing the errors, \texttt{BLOCK FnameERR}, can be defined. 


\subsection*{\texttt{BLOCK FOBSSM}}

The block \texttt{BLOCK FOBSSM} contains the SM values of the flavour
observables in the same format as in 
\texttt{BLOCK FOBS}. These SM values may be helpful as a
reference for comparison. 


\subsection*{\texttt{BLOCK FDIPOLE}}
 
The block \texttt{BLOCK FDIPOLE} contains the predictions for the electric and magnetic
dipole moments. The standard for each line in the block should correspond to
the FORTRAN format  
\begin{center} 
\texttt{(1x,I10,3x,I1,3x,I1,3x,1P,E16.8,0P,3x,'\#',1x,A)},
\end{center} 
where the first ten-digit integer should be the PDG code of a particle, the
second integer the type (electric or magnetic), the next integer the model
(SM, NP or SM+NP, as in \texttt{FWCOEF}) and finally the last double precision
number the value of the moment. The electric dipole moments must be given in
e.cm unit. 

The PDG codes for the nuclei follows the PDG particle numbering scheme
\cite{Nakamura:2010zzi}: ``Nuclear codes are given as 10-digit numbers
$\pm10LZZZAAAI$. For a (hyper)nucleus consisting of $n_p$ protons, $n_n$
neutrons and $n_\Lambda$ $\Lambda$'s, $A = n_p + n_n + n_\Lambda$ gives the
total baryon number, $Z = n_p$ the total charge and $L = n_\Lambda$ the total
number of strange quarks. $I$ gives the isomer level, with $I = 0$
corresponding to the ground state (...). To avoid ambiguities, nuclear codes
should not be applied to a single hadron, like $p$, $n$ or $\Lambda^0$, where
quark-contents-based codes already exist.'' As an example, the PDG code of the
deuteron is 1000010020.  

The types of the moments are defined as follows:\\[2mm]
\numentry{1}{electric}
\numentry{2}{magnetic}

The electric and magnetic moments can be given for the SM only, the New
Physics (NP) only or the total contributions of both SM and NP, using the
following definitions:\\[2mm] 
\numentry{0}{SM}
\numentry{1}{NP}
\numentry{2}{SM+NP}

\vspace*{-6mm}
\subsection*{\texttt{BLOCK FPARAM}}

The block \texttt{BLOCK FPARAM} contains process dependent variables, such as form factors, shape functions etc., for a specific decay.
The decay should be defined as in \texttt{BLOCK FOBS}, but with the type of the observable replaced by a user defined number for the parameter. Here it is essential to describe the variable as a comment.


\section{Conclusion}

The interplay between collider and flavour physics is entering a new era with
the start-up of the LHC. In the future more and more programs will need to be
interfaced in order to exploit maximal information from both
collider and flavour data. Towards this end, an accord will play a
crucial role. The accord presented here specifies a set of conventions
in ASCII file format for the most commonly investigated flavour-related
observables and provides a universal framework for interfacing different
programs. 

The presented accord will be further developed according to the user feedbacks and special needs of specific codes. Some discussions in this direction were held during Les Houches 2011 workshop \cite{leshouches11}, which led to a few clarifications and additions (included in the current manuscript).

As the number of flavour related codes keeps growing, the
connection between results from flavour physics and high $p_T$ physics is
becoming more relevant to the disentangling of the underlying physics model. Using the lessons learnt from the SLHA, we hope the FLHA will prove useful for studies related to flavour physics.


\subsection*{Acknowledgements}

We thank the organisers of the Les Houches 2009 workshop, where
 this work was started, for their hospitality and support. Thanks also to all those who helped us improving the accord by their useful comments, and in particular to G. B\'elanger, A. Belyaev, F. Boudjema, J.S. Lee and A. Pukhov.
This work was supported in part by the European Community's Marie Curie Research
Training Networks under contracts MRTN-CT-2006-035505
`Tools and Precision Calculations for Physics Discoveries at Colliders', and MRTN-CT-2006-035606 `MCnet'.
The work of S.H.\ was partially supported by CICYT (grant FPA 2007--66387).
The work of T.G.\ is supported in part by the Grant-in-Aid for Science
Research, Japan Society for the Promotion of Science, No.\ 20244037.

\newpage

\appendix 
\section{The PDG Particle Numbering Scheme \label{app:pdg}}

Listed in the table below are the PDG codes for some important SM mesons. Codes for other particles may be found in
\cite{Nakamura:2010zzi}. \\[1cm] 
\begin{table}[!h]
\vspace{-2ex}
\begin{center}              
\begin{tabular}{|c|c||c|c|}
\hline
Name & PDG code & Name & PDG code \\
\hline
    $\pi^0$ & 111 & $D^+$ & 411  \\
    $\pi^+$ & 211 &  $D^0$ & 421 \\
    $\rho(770)^0$ & 113 &  $D_s^+$ & 431 \\
    $\rho(770)^+$ & 213 & $D_s^{*+}$ & 433  \\
    $\eta$ & 221 & $B^0$ & 511 \\  
    $\eta^\prime(958)$& 331 & $B^+$ & 521 \\
    $\omega(782)$ & 223  & $B^{*0}$ & 513 \\ 
    $\phi(1020)$  & 333 & $B^{*+}$ & 523 \\ 
    $K_L^0$ & 130 &  $B_s^0$ & 531 \\ 
    $K_S^0$ & 310 &  $B_s^{*0}$ & 533 \\ 
    $K^0$ & 311 &  $B_c^+$ & 541 \\
    $K^+$ & 321 &  $B_c^{*+}$ & 543 \\
    $K^{*0}(892)$ & 313 & $J/\psi(1S)$ & 443 \\
    $K^{*+}(892)$ & 323 & $\Upsilon(1S)$ & 553 \\
    $\eta_c(1S)$ & 441 & $\eta_b(1S)$ & 551 \\
\hline
\end{tabular}
\caption{PDG codes for most commonly considered mesons.}
\end{center} 
\end{table}


\section{Two-Higgs Doublet Model \label{app:2hdm}}

The charged Higgs boson couplings to fermions for the Two-Higgs Doublet Model (THDM) can be expressed as

\begin{equation}
\begin{aligned}
H^+D\bar{U}:\qquad
&\frac{-g}{2\sqrt{2}m_W}V_{UD}\left[\lambda^Um_U\left(1-\gamma^5\right)
-\lambda^Dm_D\left(1+\gamma^5\right)\right]\; ,\\
H^+\ell^-\bar{\nu}_\ell:\qquad 
&\frac{g}{2\sqrt{2}m_W}\lambda^\ell m_\ell\left(1+\gamma^5\right)\; ,
\end{aligned}
\end{equation}%
where $U$, $D$ and $\ell$ stand, respectively, for the up-type quarks, the down-type quarks and the leptons. The conventions used for the four types of $Z_2$-symmetric THDM types, corresponding to different Yukawa couplings, are given in Table~\ref{tab:yukawas}.

\begin{table}[!t] 
\centering
\begin{tabular*}{0.7\columnwidth}{@{\extracolsep{\fill}}cccc}
\hline
Type &  $\lambda^U$ & $\lambda^D$ & $\lambda^\ell$ \\
\hline
I & $\cot\beta$ & $\cot\beta$ & $\cot\beta$ \\
II & $\cot\beta$ & $-\tan\beta$ & $-\tan\beta$ \\
III & $\cot\beta$ & $-\tan\beta$ & $\cot\beta$ \\
IV & $\cot\beta$ & $\cot\beta$ & $-\tan\beta$ \\
\hline
\end{tabular*}
\caption{Charged Higgs Yukawa coupling coefficients $\lambda^f$ in the 
$Z_2$-symmetric types of the THDM. \label{tab:yukawas}}
\end{table}%

The notation and meaning of the different types of model vary in the literature. Sometimes type Y (III) and type X
(IV) are used. In supersymmetry, type III usually refers to the general model encountered when the $Z_2$ symmetry
of the tree-level type II model is broken by higher order corrections.


\section{Effective Operators}
\label{app:operators}

Here we give a list of the most relevant effective operators together
with their unique two-number identifier.


\subsection{Effective operators for $b\to s$ transition}

Effective operators relevant to the $b\to s$ transition are\\
\begin{align}
O_1 &= O(03040405~6161)= (\bar{s} \gamma_{\mu} T^a P_L c)
       (\bar{c} \gamma^{\mu} T^a P_L b)\; ,\nonumber\\[4mm]
O_2 &= O(03040405~4141)= (\bar{s} \gamma_{\mu} P_L c)
       (\bar{c} \gamma^{\mu} P_L b)\; ,\nonumber\\[3mm]
O_3 &= O(03050707~4133)= (\bar{s} \gamma_{\mu} P_L b) 
       {\displaystyle\sum_q} (\bar{q} \gamma^{\mu} q)\; ,\nonumber\\[1mm]     
O_4 &= O(03050707~6153)= (\bar{s} \gamma_{\mu} T^a P_L b) 
       {\displaystyle\sum_q} (\bar{q} \gamma^{\mu} T^a q)\; ,\nonumber\\[1mm]
O_5 &= O(03050707~4536)= (\bar{s} \gamma_{\mu_1}\gamma_{\mu_2}\gamma_{\mu_3} P_L b) 
       {\displaystyle\sum_q} (\bar{q} \gamma^{\mu_1}\gamma^{\mu_2}
                                      \gamma^{\mu_3} q)\; ,\\[1mm]
O_6 &= O(03050707~6556)= (\bar{s} \gamma_{\mu_1}\gamma_{\mu_2}\gamma_{\mu_3} T^a P_L b) 
       {\displaystyle\sum_q} (\bar{q} \gamma^{\mu_1}\gamma^{\mu_2}
                                      \gamma^{\mu_3} T^a q)\; ,\nonumber\\[1mm]
O_7 &= O(0305~4422)= \dfrac{e}{16\pi^2} \left[ \bar{s} \sigma^{\mu \nu} 
       (m_b P_R) b \right] F_{\mu \nu}\; ,\nonumber\\[2mm]
O_8 &= O(0305~6421)= \dfrac{g}{16\pi^2} \left[ \bar{s} \sigma^{\mu \nu} 
       (m_b P_R) T^a b \right] G_{\mu \nu}^a\; .\nonumber
\end{align}


\subsection{Effective operators for neutral meson mixings}

Effective operators for $B_d^0-\bar{B}_d^0$ mixing are
\begin{align}
  O(01050105~4141) &=
  (\bar{d} \gamma_\mu P_L b)
  (\bar{d} \gamma^\mu P_L b)\; ,
\nonumber\\[2mm]
  O(01050105~3131) &=
  (\bar{d} P_L b)
  (\bar{d} P_L b)\; ,
\nonumber\\[2mm]
  O(01050105~7171) &=
  (\bar{d}_i P_L b^j)
  (\bar{d}_j P_L b^i)\; ,
\nonumber\\[2mm]
  O(01050105~3132) &=
  (\bar{d} P_L b)
  (\bar{d} P_R b)\; ,
\nonumber\\[2mm]
  O(01050105~7172) &=
  (\bar{d}_i P_L b^j)
  (\bar{d}_j P_R b^i)\; ,
\end{align}
and those with opposite chiralities.
Operators for  $B_s^0-\bar{B}_s^0$, $K^0-\bar{K}^0$ and
$D^0-\bar{D}^0$ mixings are defined in the same way.


\subsection{Effective operators for lepton flavour violations}

The effective operators for $\Delta LF=1$ lepton flavour
violating processes are as follows,\\
\begin{itemize}
\item
$\mu\to e$ transitions
($\mu\to e\,\gamma$, $\mu\to e\,e\,e$ and $\mu-e$ conversion in a muonic atom):\\
\begin{align}
  O(1311~4322) &=
  m_\mu (\bar{\mu} \sigma^{\mu\nu} P_L e)
  F_{\mu\nu}\; ,
\nonumber\\[5mm]
  O(13111111~3131) &=
  (\bar{\mu} P_L e)
  (\bar{e} P_L e)\; ,
\nonumber\\[2mm]
  O(13111111~4141) &=
  (\bar{\mu} \gamma_{\mu} P_L e)
  (\bar{e} \gamma^{\mu} P_L e)\; ,
\nonumber\\[2mm]
  O(13111111~4142) &=
  (\bar{\mu} \gamma_{\mu} P_L e)
  (\bar{e} \gamma^{\mu} P_R e)\; ,
\nonumber\\[5mm]
  O(13110101~3131) &=
  (\bar{\mu} P_L e)
  (\bar{d} P_L d)\; ,
\nonumber\\[2mm]
  O(13110101~3132) &=
  (\bar{\mu} P_L e)
  (\bar{d} P_R d)\; ,
\nonumber\\[2mm]
  O(13110101~4141) &=
  (\bar{\mu} \gamma_\mu P_L e)
  (\bar{d} \gamma^\mu P_L d)\; ,
\nonumber\\[2mm]
  O(13110101~4142) &=
  (\bar{\mu} \gamma_\mu P_L e)
  (\bar{d} \gamma^\mu P_R d)\; ,
\nonumber\\[2mm]
  O(13110101~4343) &=
  (\bar{\mu} \sigma^{\mu\nu} P_L e)
  (\bar{d} \sigma_{\mu\nu} P_L d)\; ,
\nonumber\\[5mm]
  O(13110202~3131) &=
  (\bar{\mu} P_L e)
  (\bar{u} P_L u)\; ,
\nonumber\\[2mm]
  O(13110202~3132) &=
  (\bar{\mu} P_L e)
  (\bar{u} P_R u)\; ,
\nonumber\\[2mm]
  O(13110202~4141) &=
  (\bar{\mu} \gamma_\mu P_L e)
  (\bar{u} \gamma^\mu P_L u)\; ,
\nonumber\\[2mm]
  O(13110202~4142) &=
  (\bar{\mu} \gamma_\mu P_L e)
  (\bar{u} \gamma^\mu P_R u)\; ,
\nonumber\\[2mm]
  O(13110202~4343) &=
  (\bar{\mu} \sigma^{\mu\nu} P_L e)
  (\bar{u} \sigma_{\mu\nu} P_L u)\; ,
\nonumber\\[5mm]
  O(13110303~3131) &=
  (\bar{\mu} P_L e)
  (\bar{s} P_L s)\; ,
\nonumber\\[2mm]
  O(13110303~3132) &=
  (\bar{\mu} P_L e)
  (\bar{s} P_R s)\; ,
\nonumber\\[2mm]
  O(13110303~4141) &=
  (\bar{\mu} \gamma_\mu P_L e)
  (\bar{s} \gamma^\mu P_L s)\; ,
\nonumber\\[2mm]
  O(13110303~4142) &=
  (\bar{\mu} \gamma_\mu P_L e)
  (\bar{s} \gamma^\mu P_R s)\; ,
\nonumber\\[2mm]
  O(13110303~4343) &=
  (\bar{\mu} \sigma^{\mu\nu} P_L e)
  (\bar{s} \sigma_{\mu\nu} P_L s)\; .
\label{eq:Op-mu-e}
\end{align}
We also define operators with opposite chiralities by replacing $P_L$
and $P_R$ with each other in (\ref{eq:Op-mu-e}).
\item
$\tau\to \mu$ transitions
($\tau\to \mu\,\gamma$, $\tau\to \mu\,\mu\,\mu$, $\tau\to \mu\, e^+\,e^-$ and $\tau\to\mu\,\mathrm{hadrons}$):
\begin{align}
  O(1513~4322) &=
  m_\tau (\bar{\tau} \sigma^{\mu\nu} P_L \mu)
  F_{\mu\nu}\; ,
\nonumber\\[5mm]
  O(15131313~3131) &=
  (\bar{\tau} P_L \mu)
  (\bar{\mu} P_L \mu)\; ,
\nonumber\\[2mm]
  O(15131313~4141) &=
  (\bar{\tau} \gamma_{\mu} P_L \mu)
  (\bar{\mu} \gamma^{\mu} P_L \mu)\; ,
\nonumber\\[2mm]
  O(15131313~4142) &=
  (\bar{\tau} \gamma_{\mu} P_L \mu)
  (\bar{\mu} \gamma^{\mu} P_R \mu)\; ,
\nonumber\\[5mm]
  O(15131111~3131) &=
  (\bar{\tau} P_L \mu)
  (\bar{e} P_L e)\; ,
\nonumber\\[2mm]
  O(15131111~3132) &=
  (\bar{\tau} P_L \mu)
  (\bar{e} P_R e)\; ,
\nonumber\\[2mm]
  O(15131111~4141) &=
  (\bar{\tau} \gamma_\mu P_L \mu)
  (\bar{e} \gamma^\mu P_L e)\; ,
\nonumber\\[2mm]
  O(15131111~4142) &=
  (\bar{\tau} \gamma_\mu P_L \mu)
  (\bar{e} \gamma^\mu P_R e)\; ,
\nonumber\\[2mm]
  O(15131111~4343) &=
  (\bar{\tau} \sigma^{\mu\nu} P_L \mu)
  (\bar{e} \sigma_{\mu\nu} P_L e)\; ,
\nonumber\\[5mm]
  O(15130101~3131) &=
  (\bar{\tau} P_L \mu)
  (\bar{d} P_L d)\; ,
\nonumber\\[2mm]
  O(15130101~3132) &=
  (\bar{\tau} P_L \mu)
  (\bar{d} P_R d)\; ,
\nonumber\\[2mm]
  O(15130101~4141) &=
  (\bar{\tau} \gamma_\mu P_L \mu)
  (\bar{d} \gamma^\mu P_L d)\; ,
\nonumber\\[2mm]
  O(15130101~4142) &=
  (\bar{\tau} \gamma_\mu P_L \mu)
  (\bar{d} \gamma^\mu P_R d)\; ,
\nonumber\\[2mm]
  O(15130101~4343) &=
  (\bar{\tau} \sigma^{\mu\nu} P_L \mu)
  (\bar{d} \sigma_{\mu\nu} P_L d)\; ,
\nonumber\\[5mm]
  O(15130202~3131) &=
  (\bar{\tau} P_L \mu)
  (\bar{u} P_L u)\; ,
\nonumber\\[2mm]
  O(15130202~3132) &=
  (\bar{\tau} P_L \mu)
  (\bar{u} P_R u)\; ,
\nonumber\\[2mm]
  O(15130202~4141) &=
  (\bar{\tau} \gamma_\mu P_L \mu)
  (\bar{u} \gamma^\mu P_L u)\; ,
\nonumber\\[2mm]
  O(15130202~4142) &=
  (\bar{\tau} \gamma_\mu P_L \mu)
  (\bar{u} \gamma^\mu P_R u)\; ,
\nonumber\\[2mm]
  O(15130202~4343) &=
  (\bar{\tau} \sigma^{\mu\nu} P_L \mu)
  (\bar{u} \sigma_{\mu\nu} P_L u)\; ,
\nonumber\\[5mm]
  O(15130303~3131) &=
  (\bar{\tau} P_L \mu)
  (\bar{s} P_L s)\; ,
\nonumber\\[2mm]
  O(15130303~3132) &=
  (\bar{\tau} P_L \mu)
  (\bar{s} P_R s)\; ,
\nonumber\\[2mm]
  O(15130303~4141) &=
  (\bar{\tau} \gamma_\mu P_L \mu)
  (\bar{s} \gamma^\mu P_L s)\; ,
\nonumber\\[2mm]
  O(15130303~4142) &=
  (\bar{\tau} \gamma_\mu P_L \mu)
  (\bar{s} \gamma^\mu P_R s)\; ,
\nonumber\\[2mm]
  O(15130303~4343) &=
  (\bar{\tau} \sigma^{\mu\nu} P_L \mu)
  (\bar{s} \sigma_{\mu\nu} P_L s)\; .
\label{eq:Op-tau-mu}
\end{align}
Operators with opposite chiralities are also defined.
\item 
We define operators for $\tau\to e$ transitions by replacing the muon
(code 13) and the electron (code 11) fields with each other in
(\ref{eq:Op-tau-mu}).
\end{itemize}

Effective operators for $\Delta LF >1$ leptonic tau decay
$\tau^+\to \mu^-\,e^+\,e^+$ are
\begin{align}
  O(15111311~3131) &=
  (\bar{\tau} P_L e)
  (\bar{\mu} P_L e)\; ,
\nonumber\\[2mm]
  O(15111311~4141) &=
  (\bar{\tau} \gamma_\mu P_L e)
  (\bar{\mu} \gamma^\mu P_L e)\; ,
\nonumber\\[2mm]
  O(15111311~4142) &=
  (\bar{\tau} \gamma_\mu P_L e)
  (\bar{\mu} \gamma^\mu P_R e)\; ,
\end{align}
and those with opposite chiralities.
Operators for $\tau^+\to e^-\,\mu^+\,\mu^+$ are defined by replacing the muon and electron fields.

For the processes in which both lepton and quark flavours are violated,
such as $\tau\to \mu\,K$, $B^0\to \mu\,\bar{e}$ and so on, relevant
operators are as follows,
\begin{align}
  O(03011513~3131) &=
  (\bar{s} P_L d)
  (\bar{\tau} P_L \mu)\; ,
\nonumber\\[2mm]
  O(03011513~3132) &=
  (\bar{s} P_L d)
  (\bar{\tau} P_R \mu)\; ,
\nonumber\\[2mm]
  O(03011513~4141) &=
  (\bar{s} \gamma_\mu P_L d)
  (\bar{\tau} \gamma^\mu P_L \mu)\; ,
\nonumber\\[2mm]
  O(03011513~4142) &=
  (\bar{s} \gamma_\mu P_L d)
  (\bar{\tau} \gamma^\mu P_R \mu)\; ,
\nonumber\\[2mm]
  O(03011513~4343) &=
  (\bar{s} \sigma^{\mu\nu} P_L d)
  (\bar{\tau} \sigma_{\mu\nu} P_L \mu)\; .
\end{align}
Definitions of operators with opposite chiralities and/or different
quark/lepton flavour combinations are straightforward.


\section{Decay constants}
\label{app:decayconst}
The decay constant $f_P$ of a pseudoscalar meson $P$ can be defined as:
\begin{equation}
  \langle 0 | \bar{q}\gamma^\mu \gamma_5 Q | P(p) \rangle = -i f_P p^\mu \;,
\end{equation}
for $q\neq Q$ quark contents ($P=\pi^\pm$, $K$, $D$, $B$).
For $\pi^0$, $\eta$ and $\eta'$, we define:
\begin{eqnarray}
  \frac{1}{\sqrt{2}}
  \langle 0 |
    \bar{u} \gamma^\mu \gamma_5 u
  - \bar{d} \gamma^\mu \gamma_5 d
  | \pi^0(p) \rangle
  &=&
  -i f_\pi p^\mu \;,
\\
  \frac{1}{\sqrt{2}}
  \langle 0 |
    \bar{u} \gamma^\mu \gamma_5 u
  + \bar{d} \gamma^\mu \gamma_5 d
  | \eta^{(\prime)}(p) \rangle
  &=&
  -i f_{\eta^{(\prime)}}^{q} p^\mu \;,
\\
  \langle 0 |
    \bar{s} \gamma^\mu \gamma_5 s
  | \eta^{(\prime)}(p) \rangle
  &=&
  -i f_{\eta^{(\prime)}}^{s} p^\mu \;,
\end{eqnarray}
assuming isospin symmetry.
Other possible choices for $\eta$ and $\eta'$ may be:
\begin{eqnarray}
  \frac{1}{\sqrt{6}}
  \langle 0 |
    \bar{u} \gamma^\mu \gamma_5 u
  + \bar{d} \gamma^\mu \gamma_5 d
  - 2 \bar{s} \gamma^\mu \gamma_5 s
  | \eta^{(\prime)}(p) \rangle
  &=&
  -i f_{\eta^{(\prime)}}^{8} p^\mu \;,
\\
  \frac{1}{\sqrt{3}}
  \langle 0 |
    \bar{u} \gamma^\mu \gamma_5 u
  + \bar{d} \gamma^\mu \gamma_5 d
  + \bar{s} \gamma^\mu \gamma_5 s
  | \eta^{(\prime)}(p) \rangle
  &=&
  -i f_{\eta^{(\prime)}}^{1} p^\mu \;.
\end{eqnarray}
In addition, the following matrix elements are defined:
\begin{eqnarray}
  (m_q + m_Q)
  \langle 0 | \bar{q} \gamma_5 Q | P(p) \rangle &=& i h_P \;,
\\
  (m_u + m_d)
  \frac{1}{\sqrt{2}}
  \langle 0 |
    \bar{u} \gamma_5 u
  - \bar{d} \gamma_5 d
  | \pi^0(p) \rangle
  &=&
  i h_\pi \;,
\\
  (m_u + m_d)
  \frac{1}{\sqrt{2}}
  \langle 0 |
    \bar{u} \gamma_5 u
  + \bar{d} \gamma_5 d
  | \eta^{(\prime)}(p) \rangle
  &=&
  i h_{\eta^{(\prime)}}^{q} \;,
\\
  2 m_s
  \langle 0 |
    \bar{s} \gamma_5 s
  | \eta^{(\prime)}(p) \rangle
  &=&
  i h_{\eta^{(\prime)}}^{s} \;.
\end{eqnarray}
The parameters $h_P$ may be unnecessary except for $\eta$ and $\eta'$ since they can be
written in terms of other quantities such as $h_\pi = m_\pi^2 f_\pi$ etc.
$h_{\eta^{(\prime)}}^{q,s}$ do not satisfy relations of this kind due to
the contributions of anomaly terms.

Decay constants of a vector meson $V$, whose quark content is $\Bar{q}Q$
(such as $\rho^\pm$ and $K^*$), are defined by the following matrix
elements,
\begin{eqnarray}
  \langle 0 | \Bar{q}\gamma^\mu Q | V(p) \rangle
  &=& m_V f_V \epsilon^\mu \;,
\\
  \langle 0 | \Bar{q} \sigma^{\mu\nu} Q | V(p) \rangle
  &=& i f^T_V ( p^\nu \epsilon^\mu - p^\mu \epsilon^\nu ) \;,
\end{eqnarray}
where $\epsilon^\mu$ is the polarisation vector of $V$.
$f_{\rho,\omega,\phi}$ in the ``ideal mixing'' limit are defined as:
\begin{eqnarray}
  \frac{1}{\sqrt{2}}
  \langle 0 |
    \Bar{u}\gamma^\mu u - \Bar{d}\gamma^\mu d
  | \rho^0(p) \rangle
&=&
  m_{\rho} f_{\rho} \epsilon^\mu \;,
\\
  \frac{1}{\sqrt{2}}
  \langle 0 |
    \Bar{u}\gamma^\mu u + \Bar{d}\gamma^\mu d
  | \omega(p) \rangle
&=&
  m_{\omega} f_{\omega} \epsilon^\mu \;,
\\
  \langle 0 |
  \Bar{s} \gamma^\mu s
  | \phi(p) \rangle
&=&
  m_{\phi} f_{\phi} \epsilon^\mu \;.
\end{eqnarray}
$f^T_{\rho,\omega,\phi}$ are also defined with the same flavour
combinations.
It is possible to define decay constants of $\omega$ and $\phi$ as
\begin{eqnarray}
  \frac{1}{\sqrt{2}}
  \langle 0 |
    \Bar{u}\gamma^\mu u + \Bar{d}\gamma^\mu d
  | \omega(\phi)(p) \rangle
&=&
  m_{\omega(\phi)} f_{\omega(\phi)}^{q} \epsilon^\mu \;,
\\
  \langle 0 |
  \Bar{s} \gamma^\mu s
  | \omega(\phi)(p) \rangle
&=&
  m_{\omega(\phi)} f_{\omega(\phi)}^{s} \epsilon^\mu \;,
\end{eqnarray}
or
\begin{eqnarray}
  \frac{1}{\sqrt{6}}
  \langle 0 |
    \Bar{u}\gamma^\mu u + \Bar{d}\gamma^\mu d
    - 2 \Bar{s}\gamma^\mu s
  | \omega(\phi)(p) \rangle
&=&
  m_{\omega(\phi)} f_{\omega(\phi)}^{8} \epsilon^\mu \;,
\\
  \frac{1}{\sqrt{3}}
  \langle 0 |
    \Bar{u}\gamma^\mu u + \Bar{d}\gamma^\mu d
    + \Bar{s} \gamma^\mu s
  | \omega(\phi)(p) \rangle
&=&
  m_{\omega(\phi)} f_{\omega(\phi)}^{1} \epsilon^\mu \;.
\end{eqnarray}


\section{Bag parameters}
\label{app:bagpar}

We define the bag parameters $B_{1,2,3,4,5}$ for $B_d^0-\bar{B}_d^0$
mixing matrix elements as follows,
\begin{align}
  \langle B_d^0 |
  (\bar{d} \gamma_\mu P_L b)
  (\bar{d} \gamma^\mu P_L b)
  | \bar{B}_d^0 \rangle &=
  \frac{2}{3} m_{B_d^0}^2 f_{B_d^0}^2 B_1 \;,
\nonumber\\
  \langle B_d^0 |
  (\bar{d} P_L b)
  (\bar{d} P_L b)
  | \bar{B}_d^0 \rangle &=
  -\frac{5}{12} m_{B_d^0}^2 f_{B_d^0}^2
  \left( \frac{m_{B_d^0}}{m_b + m_d} \right)^2 B_2 \;,
\nonumber\\
  \langle B_d^0 |
  (\bar{d}_i P_L b^j)
  (\bar{d}_j P_L b^i)
  | \bar{B}_d^0 \rangle &=
  \frac{1}{12} m_{B_d^0}^2 f_{B_d^0}^2
  \left( \frac{m_{B_d^0}}{m_b + m_d} \right)^2 B_3 \;,
\nonumber\\
  \langle B_d^0 |
  (\bar{d} P_L b)
  (\bar{d} P_R b)
  | \bar{B}_d^0 \rangle &=
  \frac{1}{2} m_{B_d^0}^2 f_{B_d^0}^2
  \left( \frac{m_{B_d^0}}{m_b + m_d} \right)^2 B_4 \;,
\nonumber\\
  \langle B_d^0 |
  (\bar{d}_i P_L b^j)
  (\bar{d}_j P_R b^i)
  | \bar{B}_d^0 \rangle &=
  \frac{1}{6} m_{B_d^0}^2 f_{B_d^0}^2
  \left( \frac{m_{B_d^0}}{m_b + m_d} \right)^2 B_5 \;.
\label{eq:bag-BdBd}
\end{align}
The renormalisation group invariant bag parameter, $\hat{B}_{B}$, can also be defined in function of $B_1$. 
Bag parameters for $B_s^0-\bar{B}_s^0$, $K^0-\bar{K}^0$ and
$D^0-\bar{D}^0$ mixings are defined in the same way.


\section{Meson mixings}
\label{app:mixing}

Meson mixing is assigned observable type 7 in \texttt{BLOCK FOBS}. In the
FLHA, we assume the standard definition for the meson mixings. 
The oscillation frequency of $Q^0_q$ and $\bar Q^0_q$ mixing is characterised by the mass difference of the heavy and light mass eigenstates~\cite{Buras:1998raa}:
\begin{equation}
\Delta M_q\equiv M_{\rm H}^{q}-M_{\rm L}^{q} = 2 {\rm Re}\, \sqrt{(M_{12}^q-\frac{i}{2}\Gamma_{12}^q)(M_{12}^{q*}-\frac{i}{2}\Gamma_{12}^{q*})},
\end{equation}
where $M_{12}$ and $\Gamma_{12}$ are the transition matrix elements from virtual and physical intermediate states respectively.
For the kaon systems this gives:
\begin{equation}
\Delta M_K = 2 {\rm Re}\, M_{12}.
\end{equation}
In the case of $\Delta B=2$ transitions since $|\Gamma_{12}| \ll |M_{12}|$, one can write:
\begin{equation}
\langle B_q^0| {\cal H}^{\Delta B=2}_{\rm eff} | \bar B_q^0\rangle = 2 M_{B_q}
M_{12}^q \,,
\end{equation}
where $M_{B_q}$ is the mass of $B_q$ meson and
\begin{equation}
\Delta M_q = 2 |M_{12}^q|.
\end{equation}
The quantity given in the block is $\Delta M_q$ and the units are fixed to
1/ps. We use the PDG number of the oscillating mesons for the parent and the
flipped sign for the daughter.%
\footnote{
Exceptions to this rule are possible, for instance, in the case of 
$K$-$\bar K$ mixing, where the states $K_{\rm long}$ and $K_{\rm short}$
with their PDG numbers 130 and 310, respectively, can be used.
}
~The number of daughters is fixed to 1. For
example, the $B_s - \bar B_s$ oscillation frequency is given as: 

\begin{verbatim}
Block FOBS  # Flavour observables
# ParentPDG type  value      q        NDA  ID1  ID2  ID3 ... comment
   531      7     1.9e01     0        1    -531 # Delta M_s
\end{verbatim}

Similarly, the corresponding SM values and the errors can be given in
\texttt{BLOCK FOBSSM} and \texttt{BLOCK FOBSERR}, respectively. 
Note that the matrix elements (which are not physical observables) cannot be
given in this block. Such quantities can be expressed in terms of Wilson
coefficients, decay constants, bag parameters, etc. which are defined in the
FLHA, or they can be given in a user defined block. 


\section{FLHA sample file}
\label{app:example}
The following is an example of a generic FLHA file. For simplicity, we have chosen a SUSY model, but instead of giving the complete SLHA spectrum, we here present it as a generic FLHA model. Therefore we use \texttt{FMODSEL} to specify the model switches rather than using \texttt{MODSEL}, which would necessitate repeating the complete SLHA spectrum. For examples of the latter, see the SLHA manuals \cite{Skands:2003cj,Allanach:2008qq}. Some lines in the blocks \ttt{FOBS} and \ttt{FOBSMS} are broken for better readability.
\bigskip
{\small
\begin{verbatim}
# SuperIso output in Flavour Les Houches Accord format
Block FCINFO  # Program information
     1     SUPERISO         # flavour calculator
     2     3.2              # version number
Block FMODSEL  # Model selection
     1     1  # Minimal supergravity (mSUGRA,CMSSM) model
     6     1  # Quark flavour is violated
Block SMINPUTS  # Standard Model inputs
     1     1.27934000e+02   # alpha_em^(-1)
     2     1.16637000e-05   # G_Fermi
     3     1.18400000e-01   # alpha_s(M_Z)
     4     9.11876000e+01   # m_{Z}(pole)
     5     4.19000000e+00   # m_{b}(m_{b})
     6     1.72900000e+02   # m_{top}(pole)
     7     1.77700000e+00   # m_{tau}(pole)
Block MINPAR   # SUSY breaking input parameters
     1     8.00000000E+02   # m_0
     2     5.00000000E+02   # m_{1/2}
     3     2.00000000E+01   # tan(beta)
     4     1.00000000E+00   # sign(mu)
     5    -5.00000000E+02   # A_0
Block FMASS  # Mass spectrum in GeV
#PDG_code  mass            scheme  scale         particle
     3     1.04000000e-01   1   2.00000000e+00   # s (MSbar)
     5     4.68474767e+00   3   0                # b (1S)
   211     1.39600000e-01   0   0                # pi+
   313     8.91700000e-01   0   0                # K*
   321     4.93700000e-01   0   0                # K+
   421     1.86960000e+00   0   0                # D0
   431     1.96847000e+00   0   0                # D_s+
   521     5.27917000e+00   0   0                # B+
   531     5.36630000e+00   0   0                # B_s
Block FLIFE  # Lifetime in sec
#PDG_code  lifetime         particle
   211     2.60330000e-08   # pi+
   321     1.23800000e-08   # K+
   431     5.00000000e-13   # D_s+
   521     1.63800000e-12   # B+
   531     1.42500000e-12   # B_s
Block FCONST  # Decay constant in GeV
#PDG_code number decay_constant scheme scale particle
   431     1   2.48000000e-01   0   0   # D_s+
   521     1   1.92800000e-01   0   0   # B+
   531     1   2.38800000e-01   0   0   # B_s
Block FCONSTRATIO  # Ratio of decay constants
#PDG_code1 code2  nb1 nb2 ratio       scheme scale comment
   321     211    1   1   1.19300000e+00   0   0   # f_K/f_pi
Block FPARAM  # Process dependent parameters
# ParentPDG number value       NDA  ID1  ID2  ID3 ... comment
     5    1   5.80000000e-01   2      3   22          # C in b->s gamma
   521    1   4.60000000e-01   3    421  -15   16     # Delta(w) in B+->D0 tau nu
   521    2   1.03000000e+00   3    421  -15   16     # G(1) in B+->D0 tau nu
   521    3   1.17000000e+00   3    421  -15   16     # rho^2 in B+->D0 tau nu
   521    1   2.68000000e-01   2    313   22          # T1 in B->K* gamma
Block FWCOEF Q= 1.60846e+02  # Wilson coefficients at scale Q
#id            order  M    value           comment
 03040405 6161   00   2    0.00000000e+00  # C1^0
 03040405 4141   00   2    1.00000000e+00  # C2^0
     0305 4422   00   2   -1.53496321e-01  # C7^0
     0305 6421   00   2   -9.51462419e-02  # C8^0
 03040405 6161   01   2    2.33177662e+01  # C1^1
 03040405 4141   01   2    0.00000000e+00  # C2^1
 03050707 6153   01   2    5.29858390e-01  # C4^1
     0305 4422   01   2   -4.27127914e-01  # C7^1
     0305 6421   01   2   -1.06024128e+00  # C8^1
 03040405 6161   02   2    3.08457152e+02  # C1^2
 03040405 4141   02   2    4.91587899e+01  # C2^2
 03050707 4133   02   2   -6.99555993e+00  # C3^2
 03050707 6153   02   2    1.25585165e+01  # C4^2
 03050707 4536   02   2    8.73907482e-01  # C5^2
 03050707 6556   02   2    1.63857653e+00  # C6^2
     0305 4422   02   2    8.40063580e-01  # C7^2
     0305 6421   02   2   -4.61747635e+00  # C8^2
Block FDIPOLE  # Electric and Magnetic dipole moments
# PDG_code  type   M   value            comment
   13         2    1   5.00716589e-10   # 1/2 (g-2)_mu
Block FOBS  # Flavour observables
# ParentPDG type  value       q           NDA   ID1   ID2   ID3 ... comment
    5    1   2.83249483e-04   0             2     3    22        # BR(b->s gamma)
  521    4   8.78354299e-02   0             2   313    22        # Delta0(B->K* gamma)
  531    1   3.44359657e-09   0             2    13   -13        # BR(B_s->mu+ mu-)
  521    1   9.89829667e-05   0             2   -15    16        # BR(B_u->tau nu)
  521    2   9.81528344e-01   0             2   -15    16        # R(B_u->tau nu)
  431    1   5.10125129e-02   0             2   -15    16        # BR(D_s->tau nu)
  431    1   5.23481907e-03   0             2   -13    14        # BR(D_s->mu nu)
  521    1   6.71755629e-03   0             3   421   -15    16  # BR(B->D0 tau nu)
  521   11   2.96312243e-01   0             3   421   -15    16  # BR(B->D0 tau nu)/
                                                                   BR(B->D0 e nu)
  321   11   6.34141764e-01   0             2   -13    14        # BR(K->mu nu)/
                                                                   BR(pi->mu nu)
  321   12   9.99918554e-01   0             2   -13    14        # R_l23
Block FOBSSM  # SM predictions for flavour observables
# ParentPDG type  value       q           NDA   ID1   ID2   ID3 ... comment
    5    1   3.10589604e-04   0             2     3    22        # BR(b->s gamma)
  521    4   8.33856375e-02   0             2   313    22        # Delta0(B->K* gamma)
  531    1   3.30981324e-09   0             2    13   -13        # BR(B_s->mu+ mu-)
  521    1   1.00845755e-04   0             2   -15    16        # BR(B_u->tau nu)
  521    2   1.00000000e+00   0             2   -15    16        # R(B_u->tau nu)
  431    1   5.10235109e-02   0             2   -15    16        # BR(D_s->tau nu)
  431    1   5.23594766e-03   0             2   -13    14        # BR(D_s->mu nu)
  521    1   6.74263430e-03   0             3   421   -15    16  # BR(B->D0 tau nu)
  521   11   2.97418437e-01   0             3   421   -15    16  # BR(B->D0 tau nu)/
                                                                   BR(B->D0 e nu)
  321   11   6.34245073e-01   0             2   -13    14        # BR(K->mu nu)/
                                                                   BR(pi->mu nu)
  321   12   1.00000000e+00   0             2   -13    14        # R_l23
\end{verbatim}
}


\end{document}